\documentclass[12pt]{article}
\usepackage{amsfonts}
\usepackage{amssymb}
\usepackage{amsmath}

\input epsf
\input{tcilatex}
\begin{document}

\thispagestyle{empty}

\begin{center}
\null\vspace{-1cm} \hfill \\[0pt]
\vspace{1cm}\medskip {\large \textbf{Heterotic $D=2 (1/3, 0)$ Susy
Models}}

\vspace{1.5cm} \textbf{M.B. Sedra} , \textbf{J. Zerouaoui} \\[0pt]
{Universit\'{e} Ibn Tofail, Facult\'{e} des Sciences,
D\'{e}partement de
Physique,\\[0pt]
Laboratoire de Physique de la Mati\`{e}re et Rayonnement (LPMR), K\'{e}%
nitra, Morocco.}\\[0pt]
\end{center}
\begin{abstract}
Following our previous work on fractional spin symmetries (FSS)
\cite{2, 4}, we build here a superspace representation of the
heterotic $D=2(1/3,0)$ superalgebra and derive a field theoretical
model invariant under this symmetry.

\end{abstract}
\newpage

\section{Introduction}

We focus here to contribute, once again, to the study of
fractional spin symmetries (FSS) \cite{1, 2, 3, 4}, a subject that
emerges remarkably in coincidence with the growing interest in
high energy and condensed matter physics through quantum field
theory \cite{5}, conformal symmetries \cite{6} and string theory
\cite{7}.

In previous works, \cite{2, 4}, we worked out the conformal field
representations of the TIM and TPM of \cite{8,9,10,11} to any
arbitrary value of the spins $s=1/k; k=2,3,...$ . This is a
particular realization showing that the $D=2$ spins $1/2$ and
$1/3$ theories are really the two leading examples of a more
general two dimensional spin $1/k$ supersymmetric theories.

Presently, this feature motivates us to develop a superspace
formulation for these kind of theories. Indeed, focusing our
attention to the $D=2 (1/3, 1/3)$, we will construct first a
superfield representation of the heterotic spin $1/3$ algebra. We
will built also a Lagrangian invariant under this symmetry. The
basic actors are the two dimensional fields of spin $0, 1/3$ and
$2/3$. There, we will show that the obtained model admits in fact
a spin $4/3$ superconformal symmetry generated  by a spin $4/3$
conserved current $G_{4/3}(z)$ in addition to the usual spin two
conformal current $T_{2}(z)$. The field realization of these
currents as well as unitarity are discussed.

\section{The Heterotic $D=2(1/3,0)$ Model}

We Start by defining the $D=2(1/3,0)$ supersymmetric algebra as
the set of operators generated by $Q_{1/3}^{-}$, $Q_{1/3}^{+}$ and
$P$ satisfying:
\begin{eqnarray}
Q^{-3} &=&P^{-} \\
Q^{+3} &=&P
\end{eqnarray}%
where we have dropped out the values of spin for reasons of
simplicity. This superalgebra is invariant under two kinds of
discrete symmetries: First the $Z_{3}$ symmetry operating as:
\begin{eqnarray}
Q^{+} &\longrightarrow &qQ^{+} \\
Q^{-} &\longrightarrow &{\bar{q}}Q^{-} \\
P &=&q^{3}P={\bar{q}}^{3}P,
\end{eqnarray}%
where $q^{3}$ = $q{^{-}}^{3}$ = 1. Second, the $Z_{2}$ symmetry,
generated by the charge conjugation operator $C$, acting on
$Q^{\pm }$ and $P$ as:
\begin{eqnarray}
CQ^{\pm }C^{-1} &=&Q^{\mp } \\
CP &=&PC
\end{eqnarray}%
Since eqs (1) and (2) are interchanged under the Z$_{2}$-symmetry,
we shall focus hereafter our attention on one equation only say
eq(1). Hermiticity of $P$ is then lost. We shall forget about this
physical requirement for the moment. Later on we will show how
this basic feature may be restored. To construct field theoretical
models exhibiting the algebra eq. (1) as a symmetry, we shall
proceed by analogy with the $D=2(1/2,0)$ supersymmetric
theory by introducing the left spin 1/3 superspace $(z,\theta ^{+})$ where $%
\theta ^{+}=\theta _{-1/3}^{+}$ is a parafermionic variable
carrying plus one $Z_{3}$-charge and a spin $s=-1/3$ and obeying:
\begin{eqnarray}
\theta ^{3} &=&0, \\
\theta ^{+}z &=&z\theta ^{+}
\end{eqnarray}%
The objects $\theta $ such that $\theta ^{k}=0,k\geq 2$ are some
how mysterious since they are not well common among the
$c-$numbers. For $k=2,$ $\theta $ is just a Grassmann variable
often used in the superspace formulation of supersymmetric
theories \cite{12}. For $k>2$, however, such objects are new in
the sense that they were not used previously. To our knowledge,
similar quantities obeying higher non linear constraints were
postulated in \cite{13}. There, they were used in the
interpretation of the Sine Gordon integrable models as a $N=2$
supersymmetric Landau-Ginszburg theory.\newline
\newline
We shall not discuss here the space of solutions of these
constraints although we shall suppose that such space exist and is
non empty. Various motivations in favor of this assumption may be
quoted. In addition to technical arguments, there are also
physical indications supporting this assumption. One of which is
the existence of models exhibiting fractional spins symmetries
\cite{14}. The latter's are generated by spin $s=\frac{1}{k}$
charge operators among which the two dimensional $s=\frac{1}{2}$
supersymmetric algebra is just the leading example. As for the
Bose-fermi local theory, two dimensional consistent fractional
spin supersymmetric theories are expected to exist and wait to be
discovered. Nevertheless, as far as the constraint equation
$\theta ^{k}=0$ is considered, solutions can be worked out. We
give hereafter two natural ones. The first solution is given by
$k\times k$ nilpotent matrices.\newline
\newline
Taking for instance $\theta $ as $\varphi \Lambda _{1}$, where $\varphi $ is
a $c$-number and where $\Lambda _{1}$ and more generally $\Lambda
_{n};-k<n<k $ are $k\times k$ matrices , expected in terms of the $%
e_{i,j},1\leq i,j\leq k$ matrix generators, having one at the site
$(i,j)$ and zero elsewhere, as $\Lambda _{n}=\Sigma _{1\leq i\leq
k}e_{i,i+n}$. These matrices obey among other properties the
identity $(\Lambda _{1})^{p}=\Lambda _{p}$ for $p<k$ and $(\Lambda
_{1})^{p}=0$ for $p\geq k$ so that $\theta ^{k}=\varphi ^{k}$
$\Lambda _{k}=0$ \cite{16}. The second
solution of $\theta ^{3}=0$\ is obtained by using two Grassmann variables $%
\psi _{1/3}^{+}$ and $\eta _{2/3}^{-}$ allowing to rewrite the equation in
the following linearized form:
\begin{eqnarray}
\psi ^{+}\eta ^{-}+\eta ^{-}\psi ^{+} &=&0 \\
\psi ^{+}\psi ^{+}+\psi ^{+}\psi ^{+} &=&2\eta ^{-}
\end{eqnarray}
where we have dropped out the index of the spin. In forthcoming
works, we shall explore this kind of solution and its relation
with $N=2$ supersymmetry. Under our hypothesis, the superspace
realization of eq.(1) generalizing the usual supersymmetric
derivative may be worked out by using covariance and dimensional
arguments. We find
\begin{eqnarray}
D^{-} &=&{{\partial }/{\partial \theta ^{+}}}+\theta ^{+2}{{\partial }/{%
\partial z}} \\
Q^{-} &=&(1+q)^{1/3}D{{^{-}}} \\
P &=&{{\partial }/{\partial z}}
\end{eqnarray}

where ${{\partial }/{\partial \theta ^{+}}}$should be understood
as a $q$ -deformed, see \cite{2} for more details. To check that
these operators form indeed a differential representation of the
algebra eq.(1), we first calculate the square of $D^{-}$. This is
a spin $2/3$ object carrying a $Z_{3}$ charge $n=1$ ($mod$ $3$).
It reads as:
\begin{equation}
{D{^{-2}=\partial }}^{2}{/{\partial \theta ^{+2}+(1+q)\theta ^{+}\partial }/{%
\partial \theta ^{+}\partial }/{\partial z+(1+q}}^{2}){{\theta ^{+2}\partial
}}^{2}{/{\partial \theta ^{+2}\partial }/{\partial z}}
\end{equation}
In deriving this relation we have used the following basic
property of the deformed \ q-derivation
\begin{equation}
{{\partial }/{\partial \theta ^{+}}}(\theta ^{+2})=(1+q){{\theta ^{+}}}%
+q^{2}\theta ^{+2}{{\partial }/{\partial \theta ^{+},}}
\end{equation}
where $q=exp(2i\pi s)$, $s=1/k,k$ positive integer. This is a
general identity valid for any value of the spin of the parafermions parameter ${{%
\theta ^{+}}}$. For a spin $s=0$ parameter ${{\theta ^{+}=x,}}$ the
deformation parameter $q$ is equal to one and then eq(16) reduces to $[{{%
\partial }/{\partial x,x}}^{2}]=2x$ or also $[{{\partial }/{\partial x,x}}%
]=1.$ For a spin $s=1/2$ parameter $q$ i.e. a Grassmann variable verifying $%
q^{2}=0,$ eq(16), reads as $[{{\partial }/{\partial \theta ,\theta
}}^{2}]=0$ since $q=-1$. An equivalent identity using the anticommutator is $\{{{%
\partial }/{\partial \theta ,\theta }}\}=1$. More generally, using the
definition of $q-$deformation of the derivative ${{\partial }/{\partial
\theta ^{+},}}$ we have on one hand
\begin{equation}
\emph{ad}{{\partial }/{\partial \theta ^{+}}}(\theta ^{+})^{n}={{\partial
\theta ^{+n}}/{\partial \theta ^{+}}}-q^{n}(\theta ^{+n}){{\partial }/{%
\partial \theta ^{+}}}
\end{equation}
and on the other hand by using the above mentioned commutation
rules:
\begin{equation}
\emph{ad}{{\partial }/{\partial \theta ^{+}}}(\theta
^{+})^{n}=(1+q+q^{2}+...+q^{n-1}){{\theta ^{+n-1}-q^{n}}}\theta ^{+n}{{%
\partial }/{\partial \theta ^{+}}}
\end{equation}
Making appropriate choices of the value of the spin, taking into
account the constraints, one recovers the known results. We turn
now to complete the proof of the equation $D^{-3}=(1+q)P$.
Repeating the same analysis, we find:
\begin{equation}
{D{^{-3}=(1+q)\partial }/{\partial z+{\partial }^{3}{/{\partial \theta ^{+3}+%
}}(1+q+q}}^{2})[(1+q){{\theta ^{+}\partial }/{\partial \theta ^{+}+{\theta
^{+2}}\partial }}^{2}{/{\partial \theta ^{+2}]\partial }/{\partial z}}
\end{equation}

The first term of the r.h.s. of this equation is the energy momentum vector
up to the coefficient $(1+q)$. The remaining terms vanish individually
either by using the identity ${{\theta ^{+3}=0}}$ or also by help of the
property $(1+q+q^{2}=0)$, the sum of all roots of $q^{M}=1$ is identically
zero.

Superfields $\phi _{r}^{n}(z,{{\theta ^{+}}})$ are superfunction defined on
the superspace $(z,{{\theta ^{+}}})$ carrying fractional values of the spin.
In our present case, the allowed spin values of $r$ are multiples of $1/3$.
Moreover, because of the nilpotency property of the variable $q$, the
superfield $\phi _{r}$ may be expanded as follows

\begin{equation}
\phi _{r}^{n}=\varphi _{r}^{n}+{{\theta ^{+}\psi }}_{r+1/3}^{n-1}+{{\theta
^{+2}\chi }}_{r+2/3}^{n-2}
\end{equation}

Under a spin $1/3$ supersymmetric infinitesimal transformation $\delta {{%
\theta ^{+}=\epsilon }}^{+}$, this superfields varies as $\delta \phi _{r}{%
=\epsilon }^{+}D^{-}\phi _{r}$. In terms of the component fields, we have:

\begin{eqnarray}
\delta \phi _{r}^{n} &=&{\epsilon }^{+}{\psi }_{r+1/3}^{n-1} \\
\delta {\psi }_{r+1/2}^{n-1} &=&{\epsilon }^{+}{\chi }_{r+2/3}^{n-2} \\
\delta {\chi }_{r+1/3}^{n+1} &=&{\epsilon }^{+}(1+q)\partial \varphi _{r}^{n}
\end{eqnarray}

where we have used the identity ${\chi }_{r+2/3}^{n-2}={\chi }_{r+2/3}^{n+1}$%
. From these equations, one learns at least two things: 1. Any
irreducible representation $\Re $ of the algebra eq.(1) is three
dimensional. The spin values of its field components are $s=0,1/3$
and $2/3(mod$ $1/3)$. Their $Z_{3}$ charges $n$ are $n=0,\pm 1$
modulo three. Note that as in spin $D=2(1/2,0)$ supersymmetric
representation theory, this ensures that the trace of spin $1/3$
supersymmetric number operator $q^{F}$ on $\Re $ vanishes
identically since

\begin{equation}
Tr(q^{F})\sim (1+q+q^{2})
\end{equation}

The second thing we would like to note is that the highest component of the
expansion of $\phi _{r}$ eq(20) transforms as a total derivative under a spin $%
1/3$ supersymmtric transformation see eqs.(21-23). Supersymmetric
invaraint actions $S$ are then defined as:

\begin{equation}
S\sim \int d^{2}zd^{2}\theta ^{+}L^{-}
\end{equation}

where the superlagrangian $L^{-}$ is required to have $1/3$ and $1$ as left
and right scale dimensions respectively. It should carry also a minus one $%
Z_{3}$ charge. As an example, we may take the non hermitean lagrangian $L^{-}
$ as:

\bigskip

\begin{equation}
L^{-}\sim (D^{-}\phi _{1}^{n})\overline{\partial }\phi _{2}^{-n},\text{ \ \
\ \ \ }n=0,\pm 1,
\end{equation}

where $\overline{\partial }=\partial /\partial \overline{z}$ and
where $\phi _{1}^{n}$ and $\phi _{2}^{-n}$ are non hermitean
superfields of charges $n$ and $-n$ respectively. Note that
superfields given by eq.(20) carry in general $Z_{3}$ charges.
This is easily seen by remarking that the representation eq. (12,
13, 14) of the algebra eq.(1) admits a $Z_{3}$ symmetry acting as
\begin{eqnarray}
{{\theta }^{+}} &\rightarrow q&{{\theta ^{+}}} \\
{z} &{\rightarrow }&q^{3}{z} \\
{{D}^{-}} &{\rightarrow }&\overline{q}D{{^{-}}} \\
{P} &{\rightarrow }&\overline{q}^{3}P
\end{eqnarray}
The $Z_{3}$-symmetry involved here is generated by the group element $%
q=exp(2i\pi /3)$. This is a discret subgroup of the continuous
$U(1)$ group of phases $U(a)=exp(ia)$; $\alpha \in \lbrack 0,2\pi
\lbrack.$\ For $\alpha =\pi ,$ one obtains the $Z_{2}$-symmetry of
the spin $1/2$ supersymmetric algebra. For $a$ arbitrary, one has
the full $U(1)$ symmetry of the $N=2U(1)$ theory \cite{16}. We
shall examine the analogy between the  $\phi _{1,3}$ deformation
of the TPM and $N=2$ supersymmetric $Z_{3}$ invaraint models
later. Using eqs.(12, 13, 14, 20) and integrating with respect to
$\theta ^{2}$, the field components Lagrangian $L$ reads as:

\begin{equation}
L\sim \partial \varphi ^{-}\overline{\partial }\overline{\varphi }^{+}+%
\overline{q}\psi _{1/3}^{+}\overline{\partial }\chi _{2/3}^{-}-q\chi
_{2/3}^{0}\overline{\partial }\overline{\psi }_{1/3}^{0}\text{ }
\end{equation}

where we have used the commutation rules $\psi ^{+}\theta
^{+}=q\theta ^{+}\psi ^{+}$ and $\chi ^{0}\theta ^{+}=q^{2}\theta
^{+}\chi ^{0}$. In this equation, we have chosen $n=-1$ as a
$Z_{3}$ charge of the superfield $\phi _{r}$ as suggested by
eq.(16). The equations of motion of the free fields are solved by

\begin{eqnarray}
\varphi  &=&\varphi (z)+\varphi (\overline{z}), \\
\overline{\psi }^{0} &=&\overline{\psi }^{0}(z), \\
\psi ^{+} &=&\psi ^{+}(z), \\
\overline{\chi }^{-} &=&\overline{\chi }^{-}(z), \\
\chi ^{0} &=&\chi ^{0}(z),
\end{eqnarray}%
\qquad \qquad In addition to its manifest $Z_{3}$-symmetry, the
above non hermitean lagrangian admits a global spin $1/3$
supersymmetric invariance generated by the conserved charge
$Q_{1/3}^{-}$

\begin{equation}
Q_{1/3}^{-}=\int dzG_{4/3}^{-}+d\overline{z}G_{-2/3}^{-}
\end{equation}

where $G_{4/3}^{-}$ and $G_{-2/3}^{-}$ are fractional spin currents
satisfying the usual conservation law namely

\begin{equation}
\overline{\partial }G_{4/3}^{-}+G_{-2/3}^{-}=0
\end{equation}

Using the transformation laws eqs.(21, 22, 23) and following the
Nother method, one may calculate explicitly these currents. We
find that $G_{-2/3}^{-}=0$ and

\begin{equation}
G_{4/3}^{-}=\partial \varphi ^{-}\overline{\psi }_{1/3}^{0}-q\psi
_{1/3}^{+}\partial \overline{\varphi }^{+}+\overline{q}\chi _{2/3}^{0}%
\overline{\chi }_{2/3}^{-}\text{ }
\end{equation}%
This is an analytic current showing that eq(31) admits indeed a
huge symmetry namely a spin $4/3$ superconformal symmetry.
Computing the variation of the lagrangian $L$ under space time
translations, we find the following field realization of the
conserved spin two tensor

\begin{eqnarray}
T_{c} &=&\partial \varphi ^{-}\partial \overline{\varphi }^{+}+\overline{q}%
\psi _{1/3}^{+}\partial \overline{\chi }_{2/3}^{-}-q\chi _{2/3}^{0}\partial
\overline{\psi }_{1/3}^{0}  \notag \\
&&-\partial (\frac{2}{3}\overline{q}\psi _{1/3}^{+}\overline{\chi }%
_{2/3}^{-}-\frac{q}{3}\chi _{2/3}^{0}\overline{\psi }_{1/3}^{0})\text{ }
\end{eqnarray}

The other components of $T_{\mu \nu }$ namely $\overline{T}$ and $\Theta $\
vanish identically as required by conformal invariance.The analyticity of $%
G^{-}$ and $T$ follows directly from eqs.(32, 33, 34, 35, 36).

Note that $T_{c}$ is a complex current, this feature was expected
since the underlying constant of motion $P=\int
(dzT_{c}+d\overline{z}\Theta )$ involved in the algebra eq.(1) is
not hermitean too. To restore this basics property, we require, to
the spin two current eq.(40), to be hemitean. This can be achieved
in two ways leading to two different theories. The first way is to
treat the degrees of freedom involved in $\phi _{1}^{-}$ and $\phi
_{2}^{+}$\ as unconstrained fields so that the hermitean energy
momentum current $T$ is equal to $T_{c}+T_{c}^{+}$ where
$T_{c}^{+}$ is the adjoint conjugate to $T_{c}$

\begin{eqnarray}
T_{c}^{+} &=&\partial \varphi ^{+}\partial \overline{\varphi }^{-}+q\psi
_{1/3}^{-}\partial \overline{\chi }_{2/3}^{+}-\overline{q}\overline{\chi }%
_{2/3}^{0}\partial \psi _{1/3}^{0}  \notag \\
&&-\partial (\frac{2}{3}q\psi _{1/3}^{-}\overline{\chi }_{2/3}^{+}-\frac{%
\overline{q}}{3}\overline{\chi }_{2/3}^{0}\psi _{1/3}^{0}),\text{ }
\end{eqnarray}

where $\varphi ^{+}=(\varphi ^{-})^{+}$ and so on. The $4/3$ supersymmetric
conserved current $G^{+}$ is obtained in a similar way. We find

\begin{equation}
G^{+}=\partial \varphi ^{+}(\overline{\psi }_{1/3}^{0})^{c}+\overline{q}\psi
_{1/3}^{-}\partial \overline{\varphi }^{-}+q(\overline{\chi }_{2/3}^{0})^{c}%
\overline{\chi }_{2/3}^{+}\text{ }
\end{equation}

where $(\overline{\psi }_{1/3}^{0})^{c}$ and $(\overline{\chi }%
_{2/3}^{0})^{c}$ are the conjugates of $(\overline{\psi }_{1/3}^{0})$ and $(%
\overline{\chi }_{2/3}^{0})$ respectively.

In this approach, one needs at least four dynamical scalar fields $\varphi
^{\pm }$ and their conjugates $\overline{\varphi }^{\pm }$. The resulting
free lagrangian reads then as

\begin{eqnarray}
L_{0} &\sim &(\partial \varphi ^{-}\overline{\partial }\overline{\varphi }%
^{+}+\partial \varphi ^{+}\partial \overline{\varphi }^{-})+(\overline{q}%
\psi _{1/3}^{+}\overline{\partial }\overline{\chi }_{2/3}^{-}+q\psi
_{1/3}^{-}\overline{\partial }\overline{\chi }_{2/3}^{+}) \\
&&\text{ -}(q\chi _{2/3}^{0}\overline{\partial }\overline{\psi }_{1/3}^{0}+%
\overline{q}\overline{\chi }_{2/3}^{0}\overline{\partial }\psi _{1/3}^{0})
\end{eqnarray}%
The second way is to require

\begin{eqnarray}
\overline{\varphi }^{+} &=&(\varphi ^{-})^{+} \\
\chi _{2/3}^{-} &=&(\chi _{1/3}^{+})^{+} \\
\chi _{2/3}^{0} &=&(\psi _{1/3}^{0})^{+}
\end{eqnarray}

relating the component fields of  $\phi _{1}^{-}$ and $\phi
_{2}^{+}$. This conjugation which affect both the charge and the
value of the spin is not new as it was introduced, although
differently, in the Zamolodchikov's $Z_{N} $ parafermionic theory
of central charge $c=2\frac{(N-1)}{(N+2)}\cite{17}$. Presently,
the eqs.(45-47) may be wondered already at the level of the
algebra (1). Indeed imposing the hermiticity of
$P=(Q_{1/3}^{-})^{3}$, one gets

\begin{equation}
(Q_{1/3}^{-})^{+}=(Q_{1/3}^{-})^{2}
\end{equation}

and vice versa. In this case the $G_{4/3}^{+}$ current does not
exist. The hermitean lagrangain $L_{0}$ under the conjugation
eqs.(45-47) coincides with the original one eq(31). \\

\end{document}